\documentclass[apl, prl, twocolumn, amsmath,amssymb]{revtex4}
\usepackage{graphicx}
\usepackage{amsfonts}
\usepackage{bm}
\usepackage{dcolumn}
\usepackage{color}

\newcommand{\hh }[1]{ \hat{\bm{#1}} }
\newcommand{\m }[1]{ \mathbf{#1} }

\newcommand{\ha }[1]{ \hat{{#1}} }

\newcommand{\mA}{\mathcal{A} }
\newcommand{\mB }{\mathcal{B} }
\newcommand{\mC }{\mathcal{C} }

\begin{document}

\title{Swimmers in thin films: from swarming to hydrodynamic instabilities}
\date{\today}

\author{Marco Leoni }
\author{Tanniemola B. Liverpool}

\affiliation{ Department of Mathematics, University of Bristol, Clifton, Bristol BS8 1TW, U.K.}

\begin{abstract}
We investigate theoretically the collective dynamics of a suspension of  low Reynolds number swimmers that are confined to two dimensions by a thin fluid film. Our model swimmer is characterized by internal degrees of freedom which locally exert active stresses (force dipoles or quadrupoles) 
on the fluid.  We find that hydrodynamic interactions  mediated by the film can give rise  to spontaneous continuous symmetry breaking (swarming), to  states with either polar or nematic homogeneous order. For dipolar swimmers, 
the stroke averaged dynamics are enough to determine the leading contributions to the collective behaviour. In contrast, for quadrupolar swimmers, our analysis shows that detailed features of the internal dynamics play an important role in determining the 
bulk behaviour. In the broken symmetry phases, we investigate fluctuations of hydrodynamic variables of the system and find that these destabilize order. Interestingly, this instability  is not \emph{generic}  and depends on length-scale. 
\end{abstract}

\maketitle

%\section{Intro}
In nature, baths of {micron-scale swimmers} are found to show remarkable out of equilibrium phenomena ranging from anomalous diffusion and viscosity enhancement, to turbulent and swirl-like motion or self organization into complex dissipative structures{~\cite{Wu, Dombrowski,   Riedel, Leptos, Sokolov} }.
They act as the inspiration for man-made devices  able to control and mix fluids on  micron scales. The search for the design principles of such devices~\cite{Dreyfus, Dhar, Gosh} remains in its  infancy.  However,  
simplified models of low Reynolds number swimmers should prove to be a useful  starting point for the theoretical understanding of this class of collective phenomena. 
 
A very simple picture{~\cite{Pedley} }then, of a swimmer is an {\em internally driven} vector oriented towards its direction of motion.
Theoretically the dynamics of a collection of such objects can be described, on long length- and time-scales, by vector  and tensor equations which are natural generalizations of equilibrium liquid crystalline hydrodynamics~\cite{degennes}. 
 These {\em active fluids} involve the study of conserved and broken symmetry variables that are non-equilibrium analogues{~\cite{Vicsek} } of Goldstone modes. Previous studies on active suspensions {~\cite{Rama, Aranson, Saintillan, Aparna} }have shown that fluctuations  in these modes, destabilize ordered states.
 {In 3d} this instability is 
  termed \emph{generic}, as it is  independent of length-scale.
However, these descriptions 
are in  fundamental sense, \emph{phenomenological} since 
 another physical mechanism~{\cite{Saintillan, Aparna} } must be invoked 
 to generate the ordered states which have subsequently been shown to be unstable{~\cite{Rama}}.

In this letter, we introduce and study a self-contained and microscopically defined physical system in which it is possible to both generate homogeneously ordered states {\em and} to examine their stability. 
This is provided by a 'suspension' of swimmers confined to two dimensions by a viscous  thin film{~\cite{Sokolov}}. 
In addition the thin film geometry is particularly accessible to experiments both from the point of view of ease of  observation and 
external activation. 
We study analytically a model that directly links the collective behaviour  to the microscopic dynamics. It combines fluctuations, both active and passive, with the deterministic motion due to activity and hydrodynamic interactions.

 We find that the purely physical coupling  mediated by the thin film can give rise to local spontaneous breaking of  symmetry  and  we identify the possible ordered states. The system can have both nematic order ;  characterized by a  macroscopic axis of mean orientation $\bf n$ and symmetry $\bf n \rightarrow - n$, or polar order ; with mean orientation axis $\bf p$ for which $\bf p \ne - \bf p$.  We then examine the stability of  each of the homogeneous phases to hydrodynamic fluctuations. We find that the isotropic phase is stable. On the other hand,  as for the bulk 3d system, the homogeneous broken symmetry states are destabilized by  the hydrodynamic modes of the system. Here, however the thin film weakens this effect and the  instabilities observed  are  length scale dependent (i.e. not {\em generic}). 
 
We restrict ourselves to a dilute solution, and we make use of a mean field approximation and study  the one-particle distribution function: $c(\m R, \hh u, t) = \sum_i\left\langle \delta ( \m R_i - \m R) \delta(\hh u_i - \hh u) \right\rangle$, the probability of finding a swimmer with average orientation $\hh u$ at position $\m R$. 
This satisfies a dynamic equation  
\begin{align}
&\partial_t c = -\nabla \cdot \m J_T  - \bm{\mathcal{R} } \cdot \bm{\mathcal{J}}_R \quad ; \label{eq:sm} \\
& \m J_T = - \m D_T \cdot  \nabla c + (\bar{v} \hh u+ \m V )c, \left. \right. \left. \right.  \bm{\mathcal{J}}_R  = - D_R \bm{\mathcal{R} } c + \bm{\Omega} c,  \nonumber
\end{align}
in terms of  translational $ \m J_T (\m R, \hh u, t)$ 
 and  rotational 
$ \bm{\mathcal{J}}_R (\m R, \hh u, t)$ currents, with $ \bm{\mathcal{R} } :=  \hh u \wedge {\partial \over \partial \hh u}$~\cite{Doi}. 
 $\m D_T$ and $D_R$ are respectively the translational (rotational) diffusion tensor (constant) of the swimmer and  represent both passive and active fluctuations.
The deterministic quantities are {\em active},  describing self-propulsion: $\bar{v} \hh u$,  and the translational $\m V$ and rotational $\bm{\Omega}$ velocities induced on a  swimmer due to the activity of others by  hydrodynamic interactions mediated by the film.

\paragraph{Thin film hydrodynamics:}  The film is  described as an infinite incompressible two dimensional layer of  fluid with (2d)  viscosity
$\eta$ filling the plane $z =0$ 
and coupled hydrodynamically to another incompressible bulk  fluid of  (3d) viscosity $\eta_e$ which
fills the region $z \neq 0$. To distinguish between them we indicate the three-dimensional quantities with a prime.
We consider  the fluid dynamics in the  vanishing Reynolds number (Stokes) limit where inertia can be neglected~\cite{Lifshitz:1987jw}. For the in-plane quantities, given an in-plane force density ${\bm F}(x,y)$  the velocity  $\bm{v}(x,y)$ and  pressure  $p(x,y)$ satisfy~{\cite{SD75} }
\begin{equation}
\eta \nabla_\perp^2 \bm{v} + \nabla_\perp p +  \sigma^+_e-\sigma^-_e = -\bm{F} \; ; \; \nabla_\perp \cdot \bm{v}=0
\label{eq:in-plane}
\end{equation}
where $\nabla_\perp=(\partial_x,\partial_y)$  is the 2d gradient operator. $\sigma^{\pm}_e := \eta_e  \partial_z\bm{v}'\vert_{0^{\pm}} $ is the shear stress of the bulk fluid at the top/bottom of the thin film;  in the external region, $z\neq 0$, the velocity  $\bm{v}'(x,y,z)$ and pressure $p'(x,y,z)$  satisfy the Stokes equation : 
\begin{math} 
\eta_e \nabla^2 \bm{v}' + \nabla p' = 0 \quad ; \quad \nabla \cdot \bm{v}'=0 \; , 
\end{math}
 where $\nabla=(\nabla_\perp,\partial_z)$  is a {3d} gradient operator.
The ratio of the two and three dimensional viscosities introduces a length-scale $s := \eta/(2
\eta_e)$ that governs two
asymptotic regimes.  For $r \ll s$ dissipation occurs almost entirely in plane and
hydrodynamic flow fields are quasi-two dimensional while at lengths $r \gg s$ dissipation is mostly due to flow out
of plane, and the hydrodynamics is similar (but {\em not} identical) to that in three dimensions

The viscous drag coefficient on a flat disk of radius $a$ embedded in the film %subject to a constant force, %(lying entirely in-plane) 
is 
\begin{math}
 \gamma = 4 \pi \eta/g\,, 
\label{eq:gamma}
\end{math}
with $g$ a function of $s/a$~\cite{SD75}. 
On length scales large compared to $a$, the interaction between several disks lying in the film can be approximated using 
point-like forces at their centers and the Green's function $\bm{H}$ of eq.~(\ref{eq:in-plane}), corresponding to the flow, $ \bm{v}(\m r) = \bm{H}(\m r - \m r_0) \cdot {\bf f}_0$ generated by a point-like force ${\bf f}_0$ at ${\bf r}_0$.
The tensor  $\bm{H}({\bf r})={s \over \eta}\int {d^2 k \over (2 \pi)^2} e^{-i {\bf k \cdot r}} {\left({\mathbb I} - \hat{\bm k} \otimes \hat{\bm k} \right) \over s  k^2 + k }$ is the thin film equivalent of the Oseen tensor~\cite{Doi}. 
In the following,  we work in the limit $r \gg s$~{\cite{, LLM, LL10} }.

We consider swimmers with an average speed $\overline{v}$ moving in a direction $\hh u$ which can be represented at large length-scales as a time-dependent force dipole that generates an associated velocity field.  We also consider swimmers for which the force dipole is zero and the leading behaviour is determined by a force quadrupole. Hence the average force density of  a swimmer with mean position,  $\m r_\alpha$ is of the form $\m f_\alpha(\m r) = -\overline{f}_d \overline{L} \hh u \ha u_i \nabla_i \delta(\m r - \m r_\alpha ) + \frac{1}{2}\overline{f}_q \overline{L}^2 \hh u \ha u_i \ha u_j \nabla_i \nabla_j \delta(\m r - \m r_\alpha ) + \cdots$, representing  the dipole and quadrupole contributions respectively where $\overline{L}$ is the typical dimension of the swimmer.

For a concrete calculation, an explicit microscopic model of a swimmer  is required and we have used  a three-disk model of swimmer~\cite{NG2004}. This is characterized by a minimum number (2) of degrees of freedom, with typical length $l$ that move in a non-reciprocal fashion in time with frequency $\omega = {2 \pi \over T}$ to achieve locomotion in the Stokes limit. In the following we indicate the time-average over a swimmer cycle period $T$ with an over-bar $\overline{h} = \frac{1}{T}\int^T_0  h(t) dt $.  
In the limit of small sinusoidal oscillations of amplitude $d$ around $l$ is possible to obtain~\cite{NG2004} the average self-propulsion velocity as $\overline{v} = \frac{a_0 \omega d^2}{6l^2} [1 +\frac{1}{\kappa^2} -\frac{1}{ (1+ \kappa)^2 }  ]$ where $a_0 := \frac{4 s}{3 g}$ and  our convention is that the swimmer move in the same direction of  $\hh u$.
For the average force we find $\overline{L} = l$ and $\overline{f}_d =  f {  d a_0 \over  12 l^2} [(\frac{1}{\kappa} -1) + 2( \frac{1}{\kappa^2} -\kappa ) + {(1 -\kappa) \over (1+ \kappa)^2 }  ]$ $\overline{f}_q =f {  d a_0 \over  36 l^2} [-1 + \frac{2}{\kappa^2} +\frac{2}{\kappa} + 2 \kappa +2 \kappa^2 +2  {\kappa \over (1+\kappa)^2}]$. Here $f := \gamma \omega d$ is the force scale  on each degree of freedom of the swimmer. The parameter $\kappa$, which is the average ratio of the internal lengths, controls the nature of the swimmer: the dipole is positive (pusher) for $\kappa >1$, negative (puller) for $\kappa <1$ and zero (quadrupole) for $\kappa =1$.  
%where $a_0 := {4 s \over 3 g}$

%\section{Two body problem}
The interaction between two such swimmers, which is the origin of  ${\bf V}, \;  \bm \Omega$ in eq. (\ref{eq:sm}), is complex  as each one is characterized by periodic internal dynamics whose cycles may also have different phases~\cite{Yeo09}. Here we will restrict ourselves to  swimmers that all have the same phase~\cite{Yeo09}, and the limit where the typical  separation $r$ between the swimmers centres $\m R^A$ and $\m R^B$ is much larger than the typical dimension  $l$ of the object. Dynamical quantities depending on hydrodynamic interactions, such as translational 
{$\m v_{2b} := \dot{\m R}^\alpha -\overline{v} \hh u^\alpha$ } and angular $\bm{\omega}_{2b} := \hh u^\alpha \wedge \dot{\hh u}^\alpha$ velocities felt by each swimmer, for $\alpha = A,B$, are described to good approximation by the first few terms of the expansion in spherical harmonics~\cite{Yeo09, Aparna} of the tensor $\bm{H}$.  
%. 
To leading order, for swimmer $A$ we find  {$\m v_{2b} =  \frac{s  \Theta^{(2)}}{2 \pi \eta}  \{  (3\mB^2-1) \frac{\hh r}{r^2} - \mB \frac{\hh u^B}{r^2} \}+ \mathcal{O}(1/r^3)$ } and  $\bm{\omega}_{2b} = \frac{s \Theta^{(2)} }{2 \pi \eta}  \{ 3  (\mA +2 \mB \mC -5 \mA \mB^2) \frac{\hh u^A \wedge \hh r}{r^3}  + (3 \mA \mB - \mC) \frac{\hh u^A \wedge \hh u^B}{r^3}  \} + \mathcal{O}(1/r^4)$.
Here $\mA := (\hh r \cdot \hh u^A)$, $\mB := (\hh r \cdot \hh u^B)$, $\mC := (\hh u^A \cdot \hh u^B)$ where  $\hh u^A$ and $\hh u^B$ are swimmers orientations. $\m r$ is the separation vector and $ \Theta^{(2)}$ is related to the time-dependent force dipole and scales as $f l$.
We neglect the effect of interactions on the internal dynamics and so do not address synchronization effects~\cite{LL10}.  
% 

%\section{Homogeneous states}
The collective dynamics is obtained by a coarse graining procedure, first in time and then in space. The velocities $\bm{\Omega},\bf V$  in  eqs.~(\ref{eq:sm})  are obtained from  the two body velocities for swimmers with positions $\m R$ and $\m R'$ and orientations $\hh u$ and $\hh u'$ as
\begin{equation}
\left( \begin{array}{c}
\bm{\Omega} \\
\m V
\end{array}\right)(\m R,  \hh u) = \int_{\hh u', \m R'}  
\left( \begin{array}{c}
\overline{\bm{\omega}}_{2b}  \\
\overline{\m v}_{2b}
\end{array}\right)(\m R-  \m R', \hh u, \hh u')
 c(\m R', \hh u')
\label{eq:m-f-vel}
\end{equation}
where, {as before}, the over-bar denotes time average over the swimmer cycle.
Order parameters such as local density $\rho$, polarization $\m P$  and nematic orientation tensor $\mathbb{S}$ are defined as moments of $c$\, :
\begin{align}
\begin{pmatrix}
\rho\\
 \m P\\
\mathbb{S} 
\end{pmatrix}(\m R, t)   = & \int_{\hh u} \begin{pmatrix}
1\\
\hh u\\
\left(\hh u \otimes \hh u - \mathbb{I}/2 \right)
\end{pmatrix} c(\m R, \hh u, t).
\label{eq:ord-param}
\end{align}

\paragraph{Homogeneous ordered states}
These are states in which $c$ and its moments defined above do not vary with position (denoted by $c^0,\rho^0,\m P^0, \mathbb{S}^0$ respectively).
%. 
Under these conditions, the mean field velocities are $\bm{\Omega}^0  = \beta_0 \hh u \wedge \m P^0 + \beta_1 \hh u \wedge \mathbb{S}^0 \cdot \hh u$  and
 $ \m V^0 = \beta_T \m P^0$
where the coefficients $\beta_0 $, $\beta_1 $ and $\beta_T$  are averaged quantities that depend on the microscopic details of  each swimmer~\cite{regular}. To leading order  $\beta_0 \approx f \frac{1 }{(16)^2 \lambda^3} \frac{s}{\eta} \frac{d}{l} (\kappa^2-1) $ and  $\beta_1 \approx  f  \frac{s}{\eta}\frac{1}{4 \lambda} \frac{a_0 d}{ l^2} \frac{1}{(1+ \kappa)^2} [ \frac{1}{3} (\kappa-1) + \frac{5}{12} (\kappa^2-\frac{1}{\kappa} ) +  \frac{1}{6} (\kappa^3- \frac{1}{\kappa^2} ) ]$.  They are both positive for $\kappa > 1$ (pushers) and negative for $\kappa < 1$ (pullers).   
%(ATT! it is fundamental the convention of swimming in the same direction of $\hh u$)
Inserting the expressions of $\bm{\Omega}^0$ and $\m V^0$ in eq~(\ref{eq:sm}) , taking the time derivative of eq~(\ref{eq:ord-param}) , 
we obtain dynamic equations for the moments of $c^0$. Density  is conserved, hence $\partial_t \rho^0 = 0$.  The others are, 
\begin{align}
 \partial_t \m P^0 = & - D_R  \m P^0  + \frac{\beta_0}{2} \rho^0 \m P^0 + \left( \frac{\beta_1}{2} -\beta_0\right) \mathbb{S}^0 \cdot \m P^0;  \label{eq:hom-ord} \\
  \partial_t \mathbb{S}^0_{ab} = & -4  D_R \mathbb{S}^0_{ab}+  \beta_0 \Big( P^0_a  P^0_b - \frac{1}{2} \delta_{ab}\big(\m P^0)^2  \Big)+ \frac{\beta_1 }{2}  \rho^0 \mathbb{S}^0_{ab} . \nonumber
\end{align}
 Analysis of eqs~(\ref{eq:hom-ord})  shows that, when $\beta_0$ or $\beta_1$  are positive,  the system can undergo a bifurcation, that signals the appearance of order. Hence pushers can give rise to order whereas pullers cannot. 
The  conditions $-D_R +\rho^0  \beta_0/2 = 0 $ and $-4 D_R + \rho^0 \beta_1/2 = 0$ define two critical lines in the space of  parameters $\rho^0$ and $f$, above which the instability occurs~\cite{Peruani08}. 
\begin{figure}[!ht]
\centering
\includegraphics[width=0.45\textwidth]{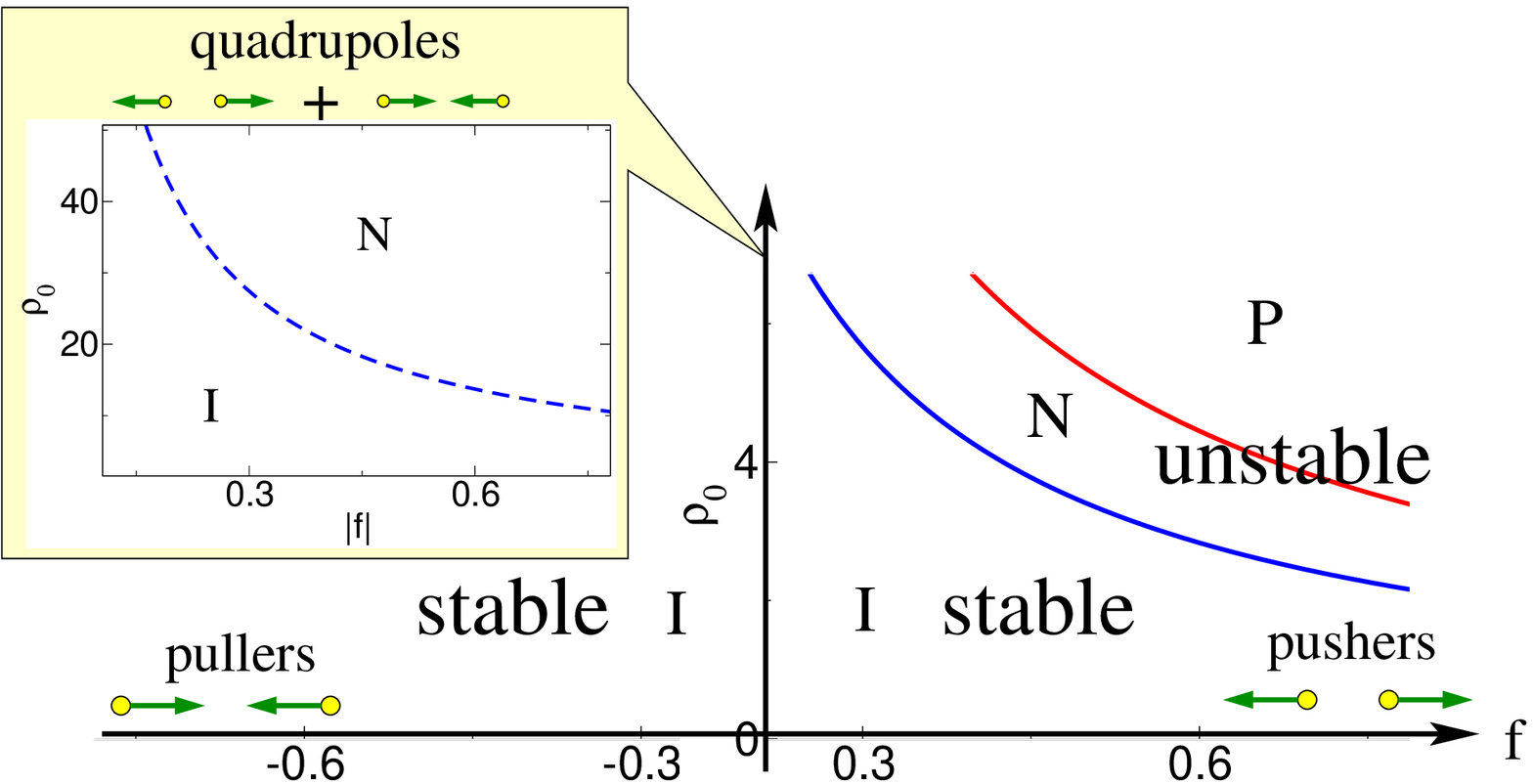}
\caption{Phase diagram showing homogeneous states and their hydrodynamic stability.  Isotropic-Nematic and Nematic-Polar transitions for dipolar swimmers. The transitions can occur only for pushers.  Force $f$ is measured in arbitrary units. Parameters are set to  $\frac{s}{\eta} = D_R = 1$, $  d = 3 a_0$, $ l = 3 d$, $\lambda =2$, and $\kappa=1.2$. { In the inset is shown  the I-N  transition for quadrupolar swimmers (same parameters  except $\kappa=1$) }. 
}  \label{fig:phase}
\end{figure}
The ratio  $ \beta_1/ (4 \beta_0)$ determines if the I-N or I-P transition occurs at lower density (see Fig. \ref{fig:phase} where we have chosen parameters for which I-N occurs first).
For pullers $\beta_0$ and $\beta_1$ are negative and  contribute to enhance the  noise (diffusion) in the system.
{For quadrupolar swimmers, these leading order terms vanish and in addition to higher order terms one must keep track of internal mode dynamics of the swimmers.  
The leading order terms in $\beta_0$ are negative, so enhance diffusion while in $\beta_1$ they are positive and promote order.  
Hence, for this particular microscopic prescription, we find no polar phase.
We find  % $\beta_0 \approx f  \frac{s}{\eta}\frac{91\pi}{4096 \lambda^3} \frac{a^2_0 d^3}{l^5}$ and 
$\beta_1 \approx f \frac{s}{\eta} \frac{a_0 d^2}{l^3} \frac{7}{384 \lambda^3} [ 1 + \frac{d}{l}(2 \pi  + 1)  ]$~\cite{LL10}. A plot of the critical line in this case is shown in the inset of Fig.~\ref{fig:phase}.  Clearly, this transition occurs  at higher densities than dipolar swimmers since it is due to higher order terms.}

%\paragraph{Fluctuations}
We now discuss fluctuations in hydrodynamic variables about isotropic and ordered states for dipole-swimmers.  As in many active systems, we find that their effect is to destabilize  order~\cite{Rama}. 
In the following we consider the deviations of the fields from their homogeneous values given by $\delta \rho = \rho - \rho^0$, similarly for $\m P$ and $\mathbb{S}$.  We introduce Fourier transforms in the usual way as $f(\m r) = \int_{\bf k} { {1 \over (2 \pi)^2} } e^{-i \m k \cdot \m r} \tilde{f}(\m k) , \tilde{f}(\m k) = \int_{\bf r} e^{i \m k \cdot \m r} f(\m r) $.
 \paragraph{Isotropic state}
In the homogeneous isotropic state $\rho = \rho^0$, $\m P^0 = \mathbb{S}^0 = 0$ and  $\rho$ is the only hydrodynamic variable.  
Variables $\delta \tilde{P}_\parallel := \delta \tilde{\m P} \cdot \hh k $, splay $\delta \tilde{\mathbb{S}}_{\parallel \parallel}$ and   bend  $(\delta \tilde{\mathbb{S}}_{\parallel \perp})_a := \delta \tilde{\mathbb{S}}_{bc} \ha k_c  (\delta_{ab} -\ha k_a \ha k_b)$ show diffusive behaviour. 
From the resulting set of equations density fluctuations are linearly stable{~\cite{Saintillan}}.

\begin{figure}[!ht]
\centering
\includegraphics[width=0.4\textwidth]{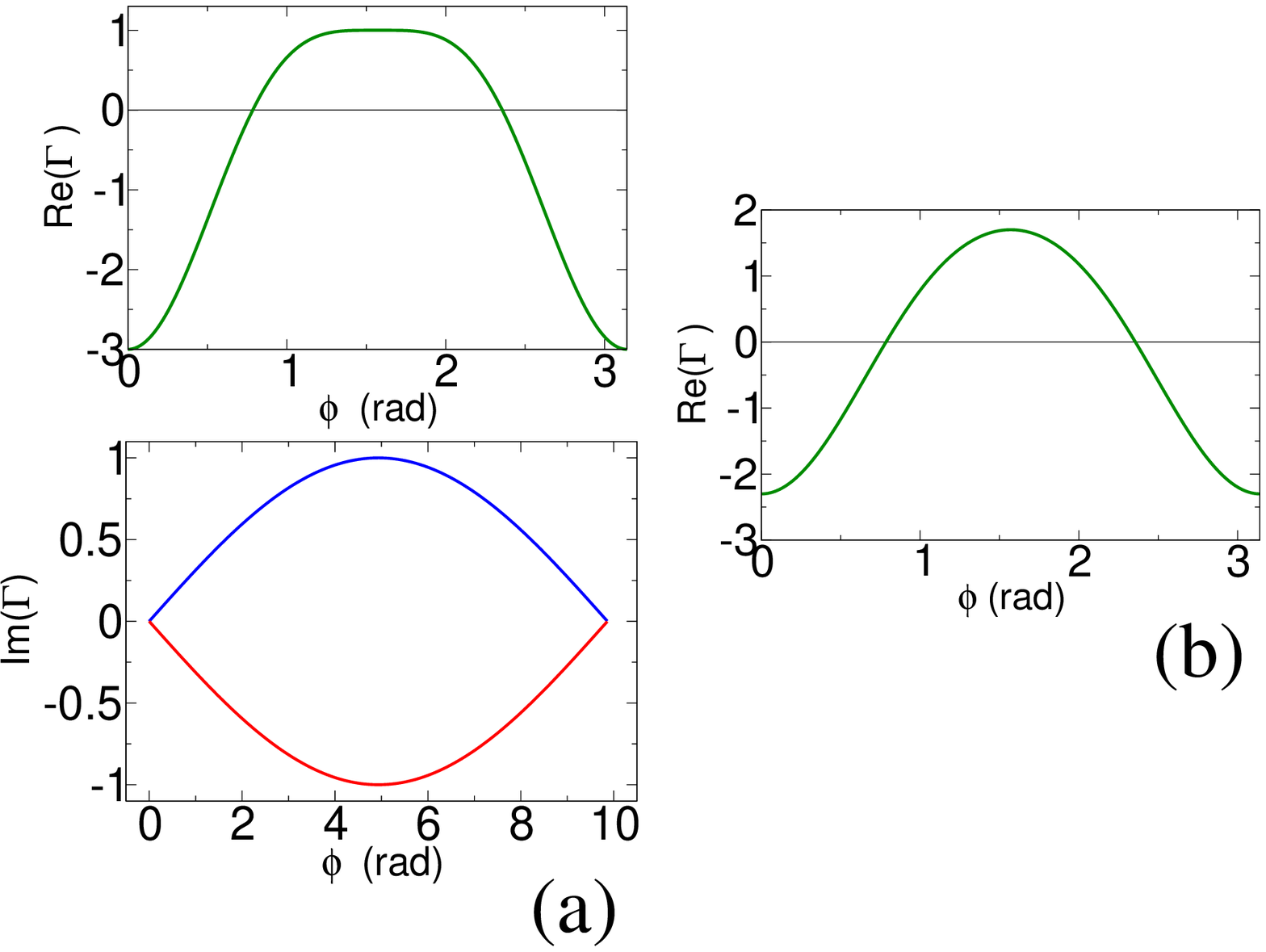}
\caption{Angular dependence of the growing mode $\Gamma$ in the splay fluctuations around homogeneous order for pushers. A positive sign indicates instability. (a) polar state. $\Re(\Gamma)$ is measured in units of $k ( s  S_P | \overline{\Theta}^{(2)} | )/(8 \eta) $ and $\Im(\Gamma)$ in units of $k \bar{v} / \sqrt{2} $; (b)  nematic state.  $\Re(\Gamma)$ is measured in units of $k (s  | \bar{\Theta}^{(2)}| )/(4 \eta)  $  and we have used  $\rho^0 = 0.3$.  }  \label{fig:instab-mode}
\end{figure}

\paragraph{Polar state}
In the homogeneous polar state $ \rho= \rho^0$, $ \m P = \m P^0$ and the hydrodynamic variables are $\rho$ and the director { $\hh P $}. The orientation tensor is slaved to $\m P$, and given by  $\mathbb{S} =  S_P P^2 ( \hh P \hh P -\frac{\mathbb{I}}{2} )$ where $S_P$ is determined by eqn. (\ref{eq:hom-ord}). The magnitude  $P$ is not a hydrodynamic variable and relaxes to a constant value on microscopic time-scales. {We set $P=1$ in the following and study  linear perturbations around this state. Decomposing $\hh k = \cos \phi \hh P^0 + \sin \phi  \hh k_\perp $, with $\hh k_\perp \cdot \hh P^0 = 0$, to leading order in $k$ we find that
%\begin{align}
%&\partial_t \delta \tilde{\rho} = 
%  i k \bar{v}  \sin \phi (\hh  k_\perp \cdot \delta \tilde{\hh P}),   \\
%& \partial_t \delta \tilde{\hh P} = 
% i \bar{v}  k\left[\frac{\hh k_\perp }{2} \sin \phi   \delta\tilde{\rho} + S_P \cos \phi \delta\tilde{\hh P} \right] + k \frac{s  S_P \overline{\Theta}^{(2)}}{4 \eta}  \nonumber \\
%&  \times \Big\{ 3  (\cos \phi)^2  \delta \tilde{\hh P} 
 %- (\sin \phi)^2  \left(1+ 4 (\cos \phi)^2 \right)   \hh k_\perp   (\hh k_\perp \cdot \delta \tilde{\hh P})   \Big\}.\nonumber 
%\end{align}
%
fluctuations in density and director are coupled via splay $(\hh k_\perp \cdot \delta \tilde{\hh P})$, }
%Defining $\frac{k_\parallel}{k} = \cos \phi$ and $\frac{k_\perp}{k} = \sin \phi$, %  
 %we find a
giving a growing mode $\Gamma$ with real part $\Re(\Gamma) \sim k  \frac{s  S_P \bar{\Theta}^{(2)}}{8 \eta}  \cos 2 \phi ( 2  + \cos 2 \phi)$  and imaginary part, {which determines the propagation speed},  $\Im(\Gamma) \sim \pm  k {\bar{v} \over \sqrt{2} } |\sin \phi |$, as shown in  figure~\ref{fig:instab-mode}(a). When $\Re(\Gamma)$ is positive fluctuations grow exponentially, signalling an instability of the ordered state due to hydrodynamic interactions, 
 analogous to those found in 3d~\cite{Rama}. However unlike those, the growth rate here scales as $k$ in the $k\rightarrow 0$ limit.

\paragraph{Nematic state}
A similar analysis can be performed around the homogeneous nematic state (finite $\rho^0$, $\mathbb{S}^0$). In this case hydrodynamic variables are density, $\rho$ and nematic orientation tensor, $\mathbb{S} = S (\hh n \otimes \hh n  -\frac{\mathbb{I}}{2} )$. Again the magnitude $S$ relaxes fast and in the following will be set to $S =1$. 
To leading order in $k$ fluctuations of density $\delta \tilde{\rho}$ and nematic director $\delta \tilde{\hh n}$ are decoupled and
%\begin{align}
%%
%& \partial_t \delta \tilde{\hh n} =  k\frac{s   \bar{\Theta}^{(2)} }{2\eta} \Bigg\{  \frac{\rho^0}{2} \Big[ \frac{k^2_\parallel}{k^2} \delta \tilde{\hh n}  + \hh k_\perp \left(1- 4 \frac{k^2_\parallel}{k^2} \right) \frac{k^2_\perp}{k^2}(\hh k_\perp \cdot \delta \tilde{\hh n} ) \Big] \nonumber \\
%& + \frac{k^2_\parallel}{k^2} \delta \tilde{\hh n} - \hh k_\perp \frac{k^2_\perp}{k^2}  (\hh k_\perp \cdot \delta \tilde{\hh n} )  
%   \Bigg\}
%\end{align}
%
%modified
%
% good
%\begin{align}
%
%& \partial_t \delta \tilde{\hh n} =  k\frac{s   \bar{\Theta}^{(2)} }{2\eta} \Big\{  \left[1+ \frac{\rho^0}{2} \right](\cos \phi)^2 \delta \tilde{\hh n}  \nonumber \\
%& + (\sin \phi)^2   \left[-1 +\frac{\rho^0}{2} \left(1-4 (\cos\phi)^2\right) \right]  \hh k_\perp  (\hh k_\perp \cdot \delta \tilde{\hh n} ) 
%   \Big\}
%\end{align}
splay fluctuations have a real mode $\Gamma \sim k  \frac{s  \bar{\Theta}^{(2)}}{4 \eta}   \cos 2 \phi (2 + \rho^0 \cos 2 \phi)$.
  For angles above $\pi/4$ destabilize the order, as shown in Fig.~\ref{fig:instab-mode} (b). 
For  quadrupolar swimmers 
% a similar analysis would show that 
the relevant terms describing hydrodynamic interactions scale as $k^2$ instead of $k$.
% {\color{blue} but do not give rise to} anti-diffusive terms. 
Their analysis is not reported here~\cite{LL10}.

It is instructive to compare our results with previous studies of swimmers in a 3d fluid~{\cite{Rama,Aparna,Saintillan}}.
Our analysis starts from a microscopic model, and `integrates out'  the fluid degrees of freedom  to see the effect on the 
other hydrodynamic modes. Alternatively one may  perform a phenomenological analysis of an ordered state 
as in ~\cite{Rama}  replacing the Stokes equation by eq~(\ref{eq:in-plane}); this  yields 
qualitatively 
 the same hydrodynamic instabilities that we have presented above. In essence, the thin film model changes the Fourier spectrum of the hydrodynamic kernel from $k^2$ to  $k + s k^2$ which in the limit  $s k \ll 1$ reduces to $\approx k$. This is the origin of the different scaling of the  instability here.
In another contrast, hydrodynamic interactions of  simplified far-field models of swimmers cannot lead to 
homogeneous order (swarming) in 3d~\cite{Aparna}. This can be ascribed to a mathematical cancellation that occurs performing  angular integrals of spherical harmonics generated by the hydrodynamic kernels. In the thin film limit considered here, this is circumvented by the confinement of the swimmer directors to two dimensions. 
In addition the instability generated by activity in the film is `soft' in the sense that it scales with $k \rightarrow 0$ 
in comparison to  the 3d case where it is independent of $k$ (hard)  to leading order~\cite{Rama}.  
We can conclude that in comparison to 3d,  hydrodynamic interactions between swimmers in a thin film favour order but are still  not strong enough to overcome the destabilizing effect of {activity}.
Finally we note that our  approach is complementary to models where simple phenomenological rules of interaction between swimmers are used to study aspects of collective behaviour~\cite{Vicsek, Chate, Peruani08}. In contrast, here we `derive' such  rules from a particular microscopic model  using a coarse-graining procedure which is valid under a precise set of conditions.

The question of how to characterise the system beyond the instability~{\cite{Chate, Saintillan} } remains open.
This highlights a fundamental theoretical issue. 
The instabilities of homogeneous ordered states seen here and in other non-equilibrium active systems 
illustrate the limits of the Landau-Ginzburg framework
~\cite{Rama,Aparna}, which has been so successful in the study { of} phase transitions in equilibrium systems. It might be that alternative approaches, such as  describing the homogeneous state as dynamical rather than stationary,  or the development of a formalism in which hydrodynamic and non-hydrodynamic variables are treated on the same level,  
have to {be} considered.

ML acknowledges the support of University of Bristol research studentship.
TBL acknowledges the support of the EPSRC under grant EP/G026440/1.

%\bibliography{notes}

\begin{thebibliography}{10}
\bibitem{regular}%
  \BibitemOpen
  \bibinfo {author} {\bibfnamefont{}~\bibnamefont{To obtain these we perform
  integrals which require a regularization at~small distances. This cut-off is naturally  provided by the swimmer size $\lambda l$, for $\lambda \ge 2$.}}%
  \bibAnnoteFile{Stop}{regular}%
\end{thebibliography}
%Merlin.mbs v4.21 2009-07-09.
%

   %\bibitem{regular}%
%To obtain these we perform integrals  which
 %require a regularization at small distances. This cut-off is naturally  provided by the swimmer size $\lambda l$, for $\lambda \ge 2$.

%comment here 

\end{document}